# Mixture Temperature-Controlled combustion: a revolutionary concept for ultra-low NO$_X$ emission

Viktor Józsa

Budapest University of Technology and Economics, Faculty of Mechanical Engineering, Department of Energy Engineering, 1111 Budapest, Műegyetem rkp. 3., Hungary

**Abstract**

Mixture Temperature-Controlled (MTC) combustion is a novel concept, offering 50% reduction in NO$_X$ emission compared to V-shaped flames without a known compromise. The essence of this concept is the central cold air injection, which is also the atomizing medium of the plain-jet airblast atomizer presently to delay fuel-air mixture ignition. Hence, the flame root is not anchored to the fuel nozzle or burner lip, facilitating distributed combustion through a homogeneous temperature field, ultimately leading to reduced NO$_X$ emission. The flame was stable up to an equivalence ratio of 0.57, which was followed by blowout as the lean flammability limit was approached. Lean combustion also means reduced flame propagation speed, being another key feature to keep the flame lifted and facilitating homogeneous mixture formation. It was observed that distributed combustion was easier to achieve under leaner conditions. Unlike flameless combustion or exhaust gas recirculation techniques, such as MILD combustion, the oxidizer can be ambient air, offering robust realization in practical applications. The distributed flame is characterized by low flame luminosity and noise. Its acoustic spectrum contains geometry-related components principally. Hence, it is hypothesized that this concept also has a lower tendency to thermoacoustic instabilities than V-shaped flames.

# 1 Introduction

Gas turbine technology for aviation and power generation is almost a century old. Initially, flame stabilization, control, and advanced materials were the focus of development [1]. The increasing concern about the environmental impact, especially at high altitudes, lead to systematic development to eliminate soot and mitigate $NO_X$ emission by the millennium [2]. This latter pollutant is of great concern also in the land since the appearance of the famous Los Angeles smog in 1943, driving even the current emission regulations towards continuously decreasing values in all related fields. Earlier combustors used non-premixed flames since they are highly stable [3], however, their pollutant emission was excessive [4].

The time scale of NO formation significantly exceeds the flow time scale [5], hence, rich burn-quick quench-lean burn (RQL) combustors were introduced [6]. The rich flame root ensures stable combustion while quick dilution makes combustion lean, leading to low $NO_X$ emission. This concept is still under development in aero engines by leading jet engine manufacturers [7], however, lean premixed prevaporized (LPP) burners offer even lower $NO_X$ emission. Nevertheless, these burners have a tendency for severe combustion instabilities [8], which is the major drawback for their general introduction in aviation. A pilot flame [9] or advanced online control algorithms can be used to solve this issue [10]. $NO_X$ and CO emission of catalytic combustion is low, nevertheless, it was never used in gas turbines due to low combustion efficiency [11]. The trivial way to achieve zero $NO_X$ emission is oxyfuel combustion, i.e., there is no $N_2$ in the oxidizer. However, the lack of an efficient $O_2$ separation unit makes this concept economically unfeasible presently [12].

To further reduce thermal $NO_X$ emission of LPP burners, exhaust gas recirculation (EGR) at high turbulence level is applied to decrease flame temperature and increase flame stability [13]. Hence, the heat release is occurring in a large volume, and pressure fluctuations are inferior to those of an LPP burner [14]. EGR implementation is not trivial in boilers [15,16] and gas turbines [17]. Nevertheless, EGR is successfully used in solid fuel combustion, e.g., in fluidized bed systems [18] and grate boilers [19]. The most widespread application is internal combustion engines, including both compression ignition [20] and spark ignition [21] variants. The technology practically building on EGR has several names around the world [13], including Moderate or Intense Low Oxygen Diffusion (MILD) combustion [17], High Temperature Air Combustion (HiTAC) [22] and Colorless Distributed Combustion (CDC) [23] to name a few. The beneficial effect of distributed combustion on $NO_X$ emission has been shown by Khalil et al. [24] and Karyeyen et al. [25] by diluting combustion air by $CO_2$ and $N_2$. The former inert gas contributed to increased CO emission, while the latter reduced it up to 18% $O_2$. Further combustion air dilution kept this pollutant unchanged.

By evaluating the temperature field of existing combustion chambers, peak suppression and high-temperature zone elimination are in the focus of development. $NO_X$ emission mitigation can be achieved by both fuel staging [26] and air staging [27,28]. The latter option includes both primary air [29] and secondary air [30] control. These techniques are all aiming to provide a more homogeneous mixture. Nevertheless, creating the perfect mixture is hindered by the realization of fuel inlet, air inlet, cooling, and other design considerations, which are necessary for reliable operation. The homogeneous mixture is also critical in hypersonic vehicles [31] and internal combustion engines [32] to meet the continuously stringent pollutant emission standards, incorporating $CO_2$ emission of cars [33].

The above concepts and their principles point to that distributed combustion in a large volume, without hot spots, would be the most beneficial solution for the mitigation of $NO_X$ emission. It also maintains a proper temperature pattern for turbine blades, allowing reduction of the share of cooling air, ultimately leading to higher efficiency [1]. A possible concept without the difficulties of EGR implementation or oxygen enrichment is Mixture Temperature-Controlled (MTC) combustion. It was demonstrated in an earlier study [34], focusing principally on various diesel-biodiesel fuel blends, that distributed combustion is possible via MTC, resulting in < 20 ppm $NO_X$ at 4.2% $O_2$. Besides flame images, spectroscopic analysis was also performed, showing very low intensity at the well-known characteristic wavelengths of hydrocarbon flames.

The present study is dedicated to the better theoretical and practical explanation of MTC combustion and distributed flame, presented in Section 2. Consequently, a reference fuel was used to evaluate operation at several conditions. The equivalence ratio was fixed in ref. [34], while it was also varied during the presented measurements. V-shaped flames were also observed at a few operating conditions in this study, hence, flame characteristics of that with distributed combustion were compared in Section 3. Lastly, combustion noise was evaluated, showing that distributed combustion via the MTC concept has a relatively low tendency to thermoacoustic instabilities.

**2 Materials and methods**

The present section begins with discussing the MTC combustion concept to highlight the practical requirements of designing a burner around it. Secondly, the experimental setup is introduced, also including some operational experiences, while the last subsection details atomization characteristics. Due to the novel concept, this section includes a more in-depth explanation than usual.

*2.1 The Mixture Temperature-Controlled combustion concept*

Steady-operating turbulent burners highly benefit from swirl vanes to ensure a homogeneous fuel-air mixture, hence, it is also a core part of the MTC combustion concept. Similar to other swirl burners, hot combustion air flows through the swirl vanes. The principal phenomenon to exploit is providing a relatively cool stream at the center to avoid the increased heat release of lean premixed burners at the flame root [35–37], which is the principal source of their NO$_X$ emission. The MTC burner solves this issue by having a central plain-jet airblast atomizer. Consequently, flashback or fuel nozzle coking cannot occur in MTC combustion due to the cold central airflow into which the fuel is injected. Besides generating a fine spray, the nozzle generates a high-speed cold air stream that surrounds the fuel stream through an adiabatic expansion as:

$$T_2 = T_1 \cdot \left(\frac{p_2}{p_1}\right)^{\frac{\gamma-1}{\gamma}}, \qquad (1)$$

where $T$ is the absolute gas temperature, $p$ is the static pressure, and $\gamma$ is the specific heat ratio of the expanding air/gas at the nozzle. Subscripts 1 and 2 represent the pre and post-expansion points. The schematic of the mentioned setup, which enables distributed combustion, is shown in Fig. 1.

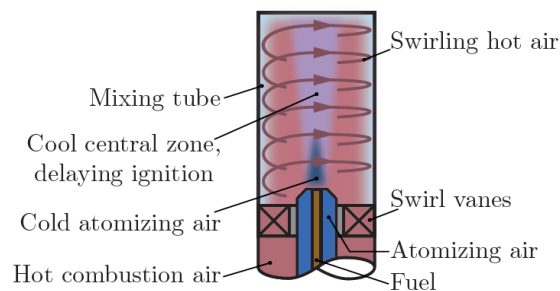

Fig. 1. Burner layout for MTC combustion.

The turbulent flow induces notable temperature fluctuations in the mixture, decreasing the heat release rate [38]. The regime diagram and the theoretical background of distributed combustion were discussed by Im et al. [39]. Also, mixture temperature correlates logarithmically with the ignition delay [40]. Both phenomena contribute to distributed combustion, facilitated by the MTC burner design. Increasing $p_1$ in Eq. (1) seems favorable, however, there is a reasonable limitation set by the application. For instance, the spray cone angle, hence, the spreading of the spray is decreasing with the increase of $p_1$ [41], even though intense turbulence facilitates spray spreading more [42]. The spray characteristics are detailed in Subsection 2.3, and the effect of expansion at the nozzle on the average mixing tube temperature is discussed in Subsection 3.1. Regardless that the present work shows the concept for airblast atomization, as the name of MTC combustion suggests, other atomizer types may be used with cold air jets to control the temperature field. A video showing distributed combustion by the presented burner is available as supplementary material of ref. [34]. This earlier work also demonstrated that flame characteristics are sensitive to fuel properties, nevertheless, the same burner allowed smooth operation with both diesel oil and 100% biodiesel. Probably, MTC burners can be designed for a wide range of fuels since fuel flexibility is increasingly important [43]. However, it has to be noted that a single design is suspected of working flawlessly in a narrower range of fuel properties.

According to the existing experiences with MTC combustion, understanding the relationship between fuel volatility and distributed combustion requires future research as there was no simple correlation observed. Diesel oil was selected for the current investigations since it is a reference fuel, however, similarly favorable characteristics were achieved with various biodiesels and lighter fuels as well. Flame luminosity is very low due to the highly homogeneous mixture in distributed combustion, similar to air dilution with inert gas [44]. Hence, the effect of radiative heat transfer has a significantly lower impact on droplet evaporation than that on fuel sprays of internal combustion engines [45].

*2.2 Experimental setup*

The experimental setup is shown in Fig. 2. The fuel was diesel oil, according to EN590:2017 standard, which was delivered from a pressurized tank. Its flow rate was measured by an Omega FPD3202 flow meter with 1.5% calibrated measurement uncertainty. The thermal power was uniformly 13.3 kW in all cases. During distributed combustion, the flame size was approximately 150×150×150 mm, meaning a 4 MW/m$^3$ volumetric heat release rate.

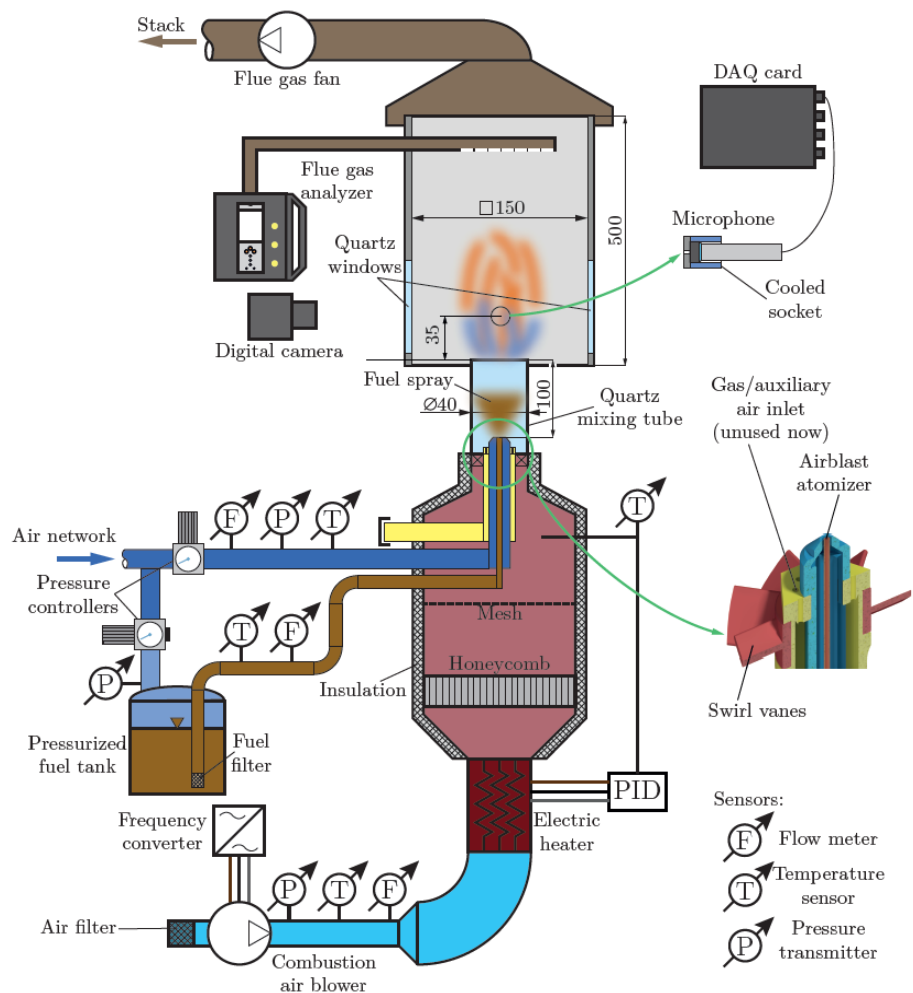

Figure 2. Principal dimensions, instrumentation, and scheme of the combustion test rig [34].

The overall equivalence ratio was varied in the range of $\phi = 0.57$–$0.86$ in four steps, corresponding to 3, 5, 7, and 9% O$_2$ concentration in the flue gas. This is the first critical parameter since NO$_X$ emission decreases with equivalence ratio, however, operating problems emerge at extremely lean conditions [1]. Stable flame for an extended time period was only possible up to 9% O$_2$; lean flame blowout was reached in a minute at 10% O$_2$, while 11% O$_2$ lead to an immediate blowout. This result is in line with the theoretical lean flammability limit of hydrocarbon fuels, which is $\phi \approx 0.5$ [46]. Besides O$_2$, CO and NO concentrations were also measured by a Testo 350 flue gas analyzer. The corresponding uncertainties are shown in Table 1, considering that the dry pollutant emissions in gas turbine applications are corrected to 15% O$_2$ [24,47]. All the discussed emission data in Subsection 3.2 were also subjected to this correction.

Table 1. Uncertainty of the flue gas analysis at 15% O$_2$.

| Gas/ $\phi$ [1] | 0.86 | 0.76 | 0.67 | 0.57 |
|---|---|---|---|---|
| NO$_X$ [ppm] | 0.669 | 0.753 | 0.861 | 1.00 |
| CO [ppm] | 1.00 | 1.13 | 1.29 | 1.50 |
| O$_2$ [V/V%] | 0.067 | 0.075 | 0.086 | 0.1 |

It was shown in Eq. (1) that the atomizing gauge pressure, $p_g$, is a governing parameter in MTC combustion. Consequently, it varied from 0.3 bar to 0.9 bar in five equidistant steps during the present measurements. Since atomizing air also contributes to $\phi$, the combustion airflow rate was hence decreased when $p_g$ was increased to balance the overall equivalence ratio. The atomizing air flow rate was measured by a pre-calibrated Omega FMA1842A flow meter with 1 liter/min uncertainty, which meant 2.3–4.2% relative uncertainty. The electric preheater provided a constant 200 °C combustion air temperature, based on previous experiences [34] since distributed combustion was observed only up to 250 °C in the case of diesel oil. All thermometers in the cold lines were B-class Pt100 resistance thermometers with 0.4 °C accuracies, while K-type thermocouples were used along the path of the hot combustion air with 2.2 °C accuracies.

The annular swirl vanes were designed for axial combustion airflow. Initially, 60° vanes were used to have a high swirl number due to the notable contribution of the axial thrust of the atomizing jet. However, the flame could not be stabilized in this case. Hence, a 45° swirl vane was used in all subsequent measurements, resulting in a geometric swirl number, $S = 0.787$

[48]. The overall swirl number, considering the momentum of the atomizing jet, is presented in Subsection 3.1.

A GRAS 146AE microphone with a DT 9837B data acquisition card was used for acoustic measurements at 20 kHz for 30 s at each setup. To keep the sensor cool, an in-house designed water-cooled socket was used with a Helmholtz resonator, tuned to 20 kHz eigenfrequency. Consequently, there is a 0–5% positive bias due to the amplification of the resonator in the spectral domain, corresponding to 0–4.3 kHz. This is acceptable in combustion since the spectral range of interest is located below this frequency [49]. A similar configuration was used by Noiray and Denisov [50] in the case of a turbulent swirl burner.

## 2.3 Atomization characteristics

Estimation of the atomization characteristics was based on the fuel flow rate at 13.3 kW thermal power and the physical properties of the fuel, shown in Table 2. The below calculations were based on a similar atomizer configuration [42]. The stoichiometric air requirement was calculated as a weighted average of that of $C_{12}H_{23}$ (14.6 kg/kg), a common diesel oil surrogate [51] and 7% biodiesel (12.4 kg/kg, based on the fatty acid composition) per the EN590:2017 standard.

Table 2. Relevant properties of the diesel oil.

| | |
|---|---|
| Lower heating value [MJ/kg] | 43 |
| Stoichiometric air requirement [kg/kg] | 14.4 |
| Fuel mass flow rate [kg/h] | 1.11 |
| Density [kg/m$^3$] | 820 |
| Surface tension [mN/m] | 25.6 |
| Kinematic viscosity [mm$^2$/s] | 2.53 |

Since the high-speed free jet also acts as a cold air stream to delay ignition, enabling distributed combustion, both the air-to-fuel mass flow ratio, *ALR*, and the momentum flux ratio,

*MFR*, are slightly higher than usual in airblast atomization. These non-dimensional quantities are defined by Eqs. (2) and (3).

$$AFR = \dot{m}_A/\dot{m}_F, \tag{2}$$

$$MFR = \rho_A \cdot w_A^2 / (\rho_F \cdot w_F^2), \tag{3}$$

where $\dot{m}$ is the mass flow rate, $\rho$ is the density, and $w$ is the flow velocity. Subscripts *A* and *F* refer to air and fuel, respectively. Since fuel evaporates from the droplet surface, the most representative droplet diameter of the generated spray in combustion is the surface-to-volume, or Sauter Mean Diameter, *SMD*. This measure of airblast atomization was found to correlate with both the Weber number, We, and the Ohnesorge number, Oh [52]. The latter one is the ratio of We and Reynolds number, Re, to cancel flow velocity. They are calculated by Eqs. (4)–(6) as:

$$\text{Re}_A = \rho_A \cdot d_0 \cdot w_A / \mu_A, \tag{4}$$

$$\text{We}_A = \rho_A \cdot d_0 \cdot w_A^2 / \sigma, \tag{5}$$

$$\text{Oh} = \sqrt{\text{We}_F}/\text{Re}_F = \mu_F/(\sigma \cdot d_0 \cdot \rho_F)^{0.5}, \tag{6}$$

where $d_0$ = 1.2 mm is the initial diameter of the liquid jet, $\mu$ is the dynamic viscosity, and $\sigma$ is the surface tension. *SMD* of the spray can be estimated by Eq. (7), based on [42]:

$$SMD/d_0 = (0.477 \cdot \text{We}_A^{-0.5} + 0.50 \cdot \text{Oh}) \cdot (1 + 1/AFR). \tag{7}$$

The above-detailed variables for all $p_g$ values are presented in Table 3, except Oh. It was 0.015 at all conditions, as it is calculated from the physical properties of the diesel oil and the initial liquid jet diameter.

Table 3. Characteristic quantities of atomization as a function of $p_g$.

| $p_g$ [bar] | AFR [1] | MFR [1] | $Re_A$ [1] | $We_A$ [1] | SMD [μm] |
|---|---|---|---|---|---|
| 0.3 | 1.55 | 87.2 | 10556 | 884 | 34.9 |
| 0.45 | 1.89 | 124 | 13149 | 1264 | 28.8 |
| 0.6 | 2.17 | 160 | 15448 | 1622 | 25.5 |
| 0.75 | 2.44 | 198 | 17755 | 2003 | 23.1 |
| 0.9 | 2.69 | 234 | 19921 | 2370 | 21.4 |

Based on Re and We, the jet breakup mechanism is atomization [53], and the droplet breakup mode is shear breakup [54].

## 3 Results and discussion

This section details the swirl number and the average inlet temperature, followed by the characteristic flame shapes at various setups. Selected flame images are also analyzed in the first subsection. Then $NO_X$ and CO emissions are shown in Subsection 3.2, quantifying the difference between the presented flames. The spectral analysis of the acoustic signal is discussed in Subsection 3.3, comparing the observed flame shapes.

### 3.1 Flame characteristics

MTC combustion features a weak swirl ($S < 0.6$), even though the swirl vane would otherwise generate a strong swirl, leading to a V-shaped flame. The axial momentum of the atomizing jet significantly reduces $S$, while the increasing preheated combustion airflow rate at lower equivalence ratio values counteract it, shown in Fig. 3a. Also, the overall combustion air temperature, $T_{OCA}$, is affected, presented in Fig. 3b. Increasing $p_g$ results in lower temperature after expansion at the atomizer nozzle, according to Eq. (1), which is counteracted by the increasing combustion airflow rate at lower $\phi$. Hence, the trend is similar to that of Fig. 3a. Figure 3c shows the percentage of the atomizing air mass flow rate and the total airflow rate, $r_A$. This is the inverse of Figs. 3a and 3b, as the high atomizing airflow rate results in low $S$ and $T_{OCA}$.

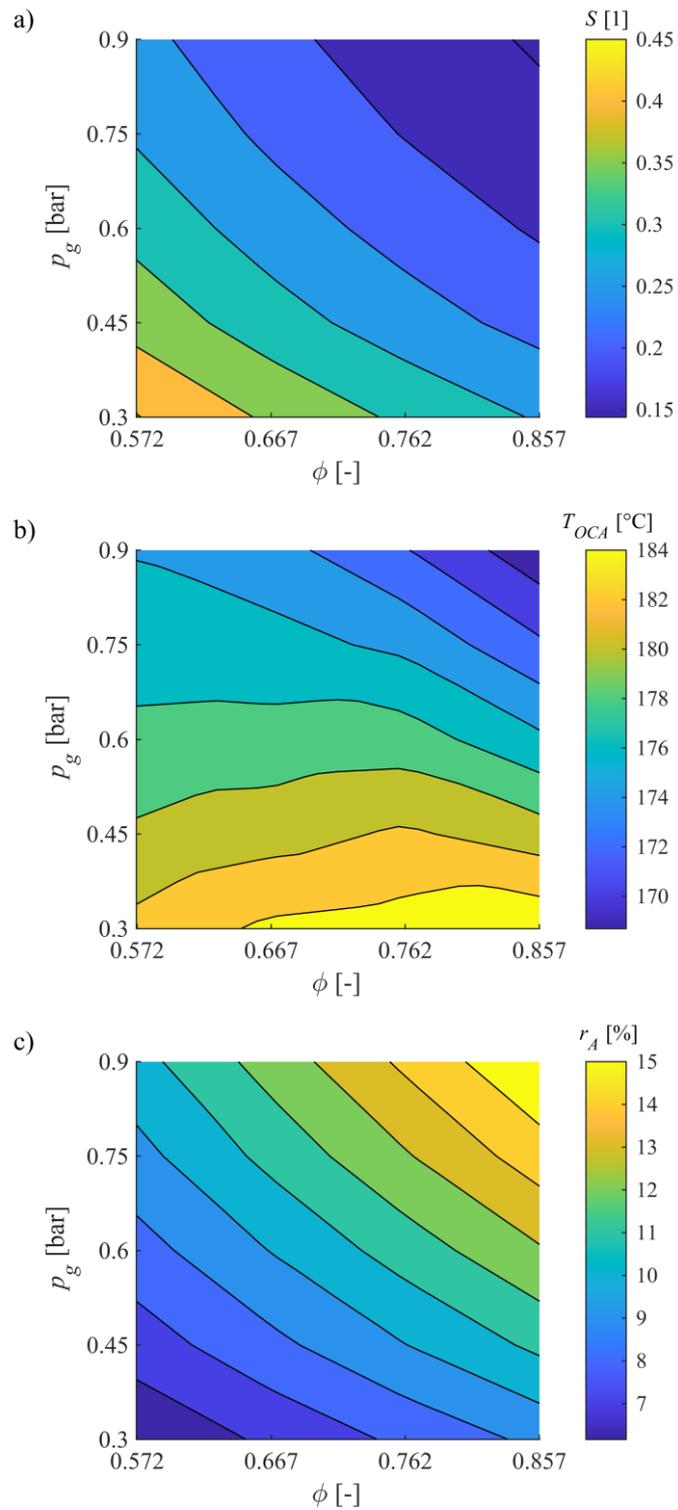

Fig. 3. a) swirl number, b) overall combustion air temperature, and c) mass flow ratio of cold atomizing air to total air mass flow rate.

There were three stable flame shapes distinguished during combustion tests. They were straight flame, V-shaped flame, and distributed combustion, corresponding to the well-set conditions of MTC burner. Also, a transitory behavior was observed between various stable flame shapes, as shown in Fig. 4.

| $p_g$ [bar] | $\Phi$ [°C] | | | |
|---|---|---|---|---|
| | 0.57 | 0.67 | 0.76 | 0.86 |
| 0.9 | d | d | d | s |
| 0.75 | d | d | d | s |
| 0.6 | d | d | d | s |
| 0.45 | d | t | t | s |
| 0.3 | v | v | s | s |

Fig. 4. Flame shapes at each measurement point. Light blue (d): distributed combustion; brown (v): V-shaped flame; orange (s): straight flame; light green (t): transitory flame between the upper and lower neighboring flame shapes.

The inhomogeneous flow field, i.e., the swirling hot air outside and the cold central axial airflow characterizes the operation of this burner setup since V-shaped flame would not be achievable otherwise at such a low $S$. Distributed combustion was only possible with leaner mixtures and from $p_g = 0.45$ bar. Consequently, the higher the temperature difference between the central region and the annular swirling flow, the more favorable the conditions are for such operation. This is facilitated by the following phenomena.

- The ignition delay time is exponentially increasing with decreasing temperature.
- Two fluids with notably different viscosity do not mix. The cold air packets, which contain fuel, need to heat up in the turbulent flow to meet the conditions of ignition. Turbulence stretches these packets, however, the majority of heat transfer occurs via conduction, which is a relatively slow process in gases.
- Lean operation means reduced fuel concentration, hence, lower flame speed and narrower parameter range for ignition [1]. These conditions are all favorable to have a lifted flame, allowing enough time for homogeneous mixture formation.

It is noteworthy that distributed combustion was not observed at the highest equivalence ratio, even though $p_g = 0.9$ bar was also tested. Consequently, having the highest share of the atomizing air and the lowest average air temperature, according to Fig. 3, does not automatically lead to more favorable conditions. This result indicates that extensive further research is required to understand the criteria of distributed combustion by the MTC concept.

The following conclusions can be drawn based on the results of the present and the preceding research [34] as a requirement of distributed combustion by using the MTC concept. Air preheating should be low since ignition occurs too early at excessive temperatures, which was about 350 °C and above for the current burner. It varies slightly with the fuel type and burner design might also influence it. The lower limitation of air preheating is determined by proper fuel vaporization. Stable diesel oil combustion was impossible at 100 °C air temperature, and thin fuel film accumulation was observed on the mixing tube wall at 150 °C. At 200 °C, all the walls were dry, indicating appropriate droplet vaporization. According to Fig. 3c, ~12% air injection via the atomizer is advised. Increasing this value and hence $p_g$ even further does not have a notable favorable impact on the flame. Figure 4 indicates that distributed combustion requires $\phi = 0.76$ or lower, where the limitation is the lean flame blowout.

Figure 5 shows six images of various stable flame shapes. The top row corresponds to straight flames, presenting the effect of atomizing pressure on the flame structure. Increasing $p_g$ results in decreasing *SMD*, hence, droplet evaporation and mixing with the combustion air occurs earlier. Consequently, flares are disappearing, flame luminosity decreases, and the fuel-air mixture becomes more homogeneous.

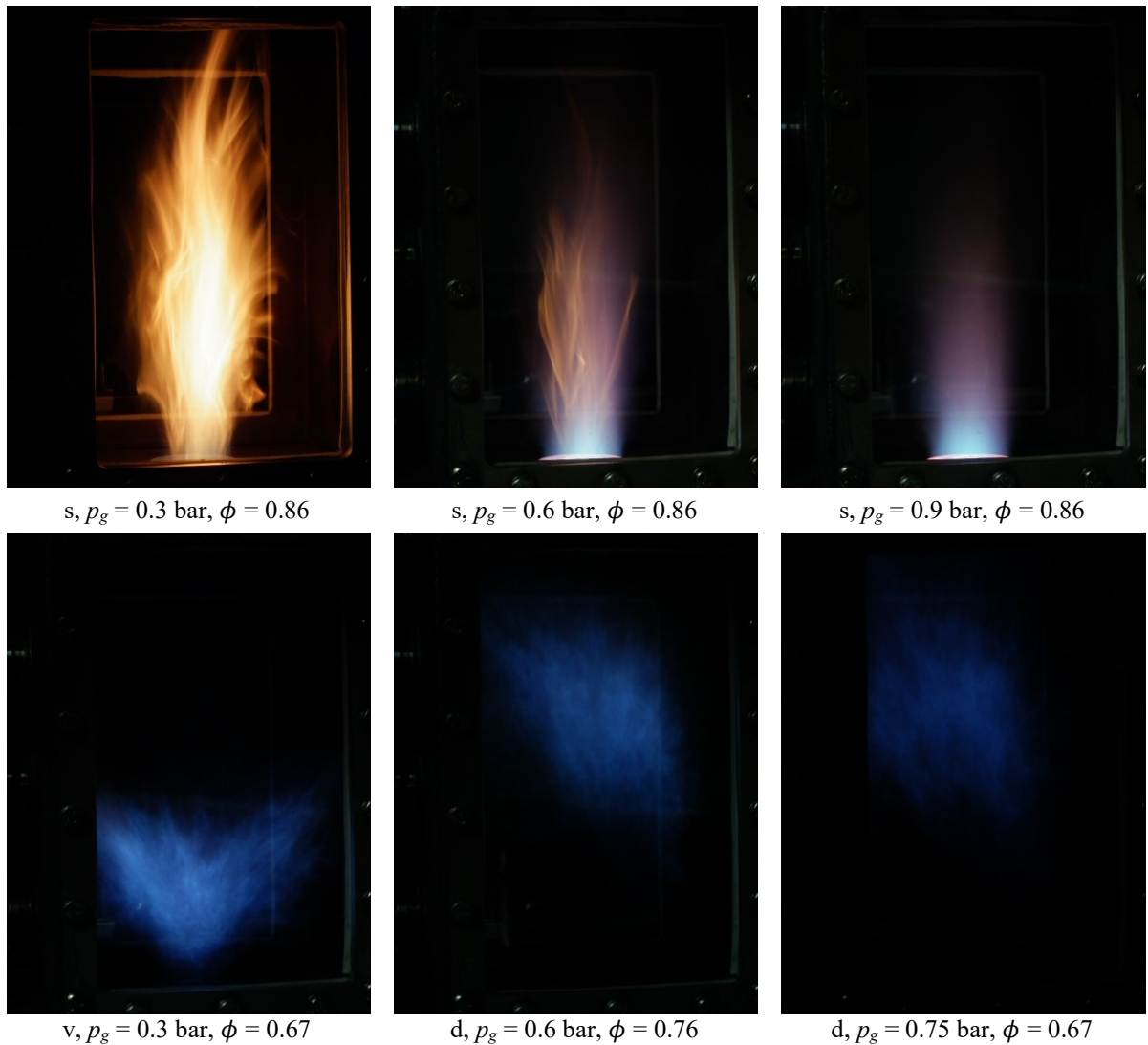

Fig. 5. Flame images at representative states. Top row: straight flames at a fixed equivalence ratio and various atomizing pressure values. Bottom row: V-shaped flame, and distributed combustion. All images were uniformly taken with 1/30 s, ISO–400, and f/4 settings.

The bottom row of Fig. 5 contains a V-shaped flame and two distributed flames. The former one has a significantly higher luminosity, however, this is far lower than that of the straight flames. In the case of distributed combustion, the presented images are brighter than the other ones captured subsequently, similar to the effect of combustion air dilution [44]. This result ultimately points to the advantage of distributed combustion: the flame occupies a large volume. Hence, heat release is less intense due to the more homogeneous mixture, qualitatively approaching flameless/MILD combustion [17]. However, combustion air does not have to be diluted with either an inert or recirculated flue gas, making this novel combustion concept highly attractive for gas turbine applications. Even more so, as the leaner the mixture, the distributed combustion is easier to achieve and maintain. Distributed combustion overcomes a disadvantage of the V-shaped flames: the heat release in the flame root occurs in a small volume, unavoidably leading inhomogeneous temperature field.

All operating points were investigated for at least one minute, while the average was three minutes. The flame of distributed combustion was well-localized, and no blowout or notable acoustic fluctuations were occurred up to $\phi = 0.57$. Consequently, it can be stated that distributed combustion matches the blowout stability of all other flame shapes observed presently.

*3.2 Pollutant emissions*

The ultimate measure of a new combustion concept from the viewpoint of regulations is the offered reduction in pollutant emissions. Nevertheless, flame stability, operational flexibility, availability, and potentials in burner tuning for increased efficiency are all critical in industrial technologies. Among the pollutants, $NO_X$ is of greatest concern since it can be avoided in the case of perfect combustion, which exists only theoretically.

Figure 6a shows the $NO_X$ emission at all conditions, corrected to 15% $O_2$ in all cases. There are two general trends. The first one is that straight flame is characterized by high $NO_X$ concentration since the released heat is concentrated to a small volume. The other one is the decreasing concentration with decreasing $\phi$ and increasing $p_g$, as both dilution and lower overall combustion air temperature decrease the adiabatic flame temperature. $NO_X$ emission of transitory flames is similar to that of the flame shape with higher emission.

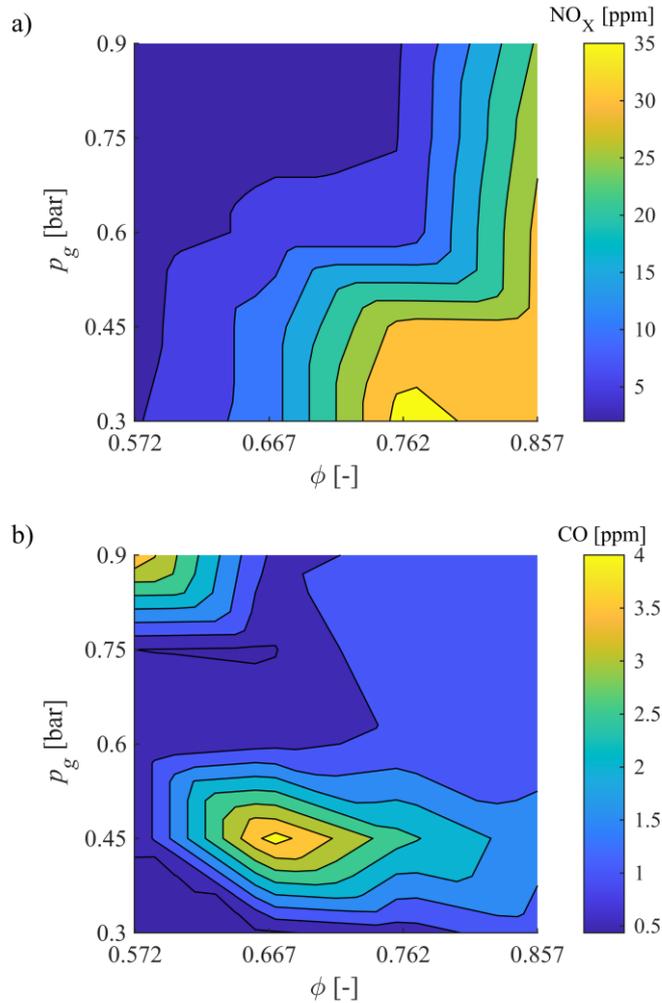

Fig. 6. Pollutant emissions: a) $NO_X$, b) CO. All data were corrected to 15% $O_2$.

Both V-shaped flames and distributed combustion are characterized by low $NO_X$ compared to the straight flame, however, the latter one features a 53% reduction on average compared to the former one. More precisely, the $NO_X$ emission of the V-shaped flame was 12.4 ppm at $\phi$= 0.67, while the average of that of distributed combustion was 4.7 ppm at the same equivalence ratio. At $\phi$= 0.57, these values were 4.5 ppm and 2.5 ppm, respectively. Nevertheless, further reduction is possible using flameless combustion techniques [17,22,24,25] since the adiabatic flame temperature will also decrease. However, neither Khalil et al. [24] nor Karyeyen et al. [25] were able to reach similarly low $NO_X$ emission by using ambient air as the oxidizer.

Figure 6b shows CO emission, which can be considered marginal at all conditions, compared to current pollutant emission limitations worldwide. Since there is no correlation between the two emissions, it can be concluded that all three flame shapes are appropriate for complete combustion. The qualitatively outstanding points are local features, hence, there is no obvious fundamental reason for them. Concluding from the pollutant emission data, MTC combustion is a highly favorable concept since it provides a further significant decrease in $NO_X$ emission compared to the widely used V-shaped flames, while the CO emission remains low.

*3.3 Acoustic characteristics*

The acoustic spectrum of various setups, which were discussed earlier in Fig. 5, is shown in Fig. 7. The Fourier transformation was performed with a 4096 sample window, using Hamming weighting. The instantaneous results are shown in Fig. 7a, indicating a high variation of the sound pressure level, *SPL*. It is caused by the temporal fluctuation of the temperature field, affecting the speed of sound, hence, the characteristic frequencies. Consequently, averaging the 4096 sample windows with a 50% overlap was performed over the 30 s signal, shown in Fig. 7b. There was zero (Z) spectral weighting used. The recorded signal of the selected conditions was free from temporal bursts, hence, spectral bias. Consequently, these results are respective to a smooth operation.

The straight flame was characterized by the highest overall *SPL*, *OASPL*, 123.5 dB. That of the background noise, i.e., without combustion, was 108.4 dB, which was originated from the shearing flow of the atomizing free jet and the external cooling jets of the glass windows. The overall noise of the V-shaped flame was 119.5 dB, notably lower than that of the straight flame. Distributed combustion resulted in the lowest *OASPL* values, 114.0 and 113.0 dB at $\phi = 0.76$ and 0.67, respectively. Consequently, this combustion mode offers a significant reduction in the overall acoustic load.

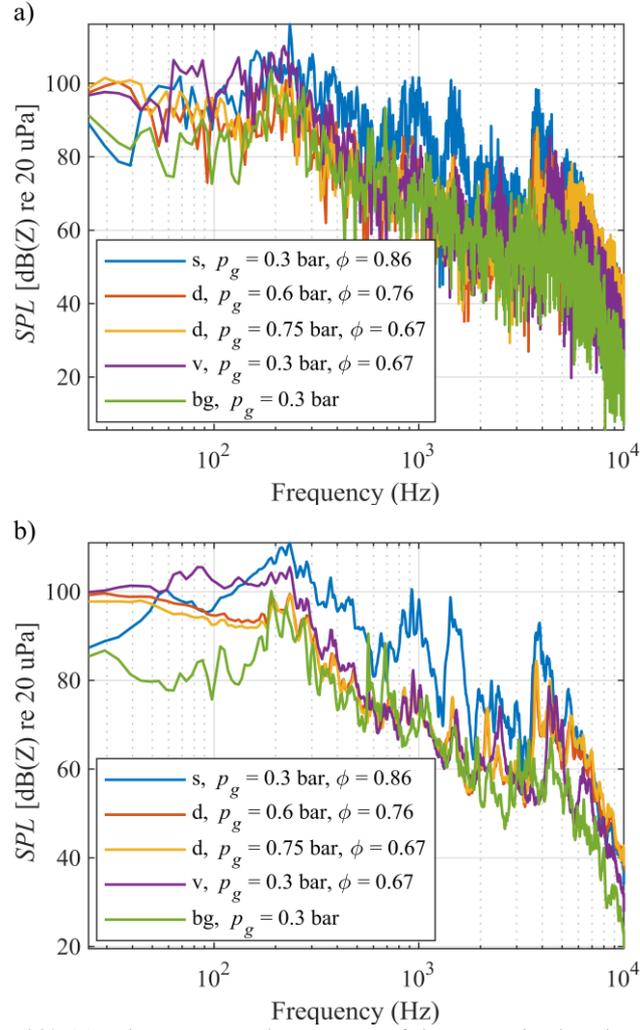

Fig. 7. a) instantaneous and b) 30 s time-averaged spectrum of the acoustic signal at various setups, related to the flame images of Fig. 5. bg denotes background noise.

The averaging used in Fig. 7b also helps in identifying the characteristic peaks, summarized in Table 4 with their hypothesized source, assuming rectangular duct and cylinder geometries [55]. This is possible due to the fact that combustion, as a phenomenon, has no characteristic frequency; the peaks are the result of the interaction of the chemical reactions with the flow field and the combustion chamber, CC, geometry. The highest *SPL* peak is suitable for determining the average speed of sound since it corresponds to the quarter-wave inside the combustion chamber, i.e., the wavelength is 0.5 m × 4 = 2 m. The highest *SPL* in the hot cases, H, was uniformly located at 234.4 Hz, implying a similar temperature field average, originated from the identical thermal power. Nevertheless, this result is not evident; both the different heat release patterns and the flame luminosity differences would suggest at least flame shape-dependent frequency peaks, which were observed at higher frequencies. The average speed of sound was hence $a_H = 468.8$ m/s. In the case of the background noise (noted as the cold case, C), the peak frequency was 190.4 Hz, meaning $a_C = 388.8$ m/s. Even though $a_H$ and $a_C$ values alone seem low in the field of combustion, it should be noted that the central cold jet inside the mixing tube, MT, significantly affects the temperature field, so the propagation of the sound waves.

Table 4. Characteristic frequencies and their sources.

| Frequency [Hz] | Case | Wave | Source |
|---|---|---|---|
| 190.4 | C | 1/4 | CC length: 0.5 m |
| 234.4 | H | 1/4 | CC length: 0.5 m |
| 566.4 | C | 3/4 | CC length: 0.5 m |
| 683.6 | C | 1/4 | MT length: 0.116 m |
| 835.0 | H | 1/2 | CC width: 0.3 m |
| 922.9 | H | 5/4 | CC length: 0.5 m |
| 1001 | H | 1/4 | MT length: 0.116 m |
| 1425 | H | 1 | CC width: 0.3 m |
| 3877 | H | 2 | CC width: 0.3 m |

Local frequency peaks apart of the listed ones in Table 4 are most probably originated from various flow structures. Revealing these sources requires computational fluid dynamic simulations, which is a potential direction for future research. The time-averaged *SPL* of distributed combustion in Fig. 7b contains primarily well-localized peaks related to the combustion chamber and burner geometry. Hence, it can be concluded that thermoacoustic instabilities are less likely to endanger stable operation of distributed combustion near the lean blowout limitation than that in the case of V-shaped flames. This favorable characteristic is similar to that of flameless and MILD combustion concepts [13].

# 4 Conclusions

- The essence of the Mixture Temperature-Controlled combustion concept is the cold central air inlet, delaying ignition. Consequently, combustion occurs downstream of the burner, occupying a large volume in the combustion chamber. Hence, this burner configuration is free from flashback and fuel nozzle coking by design.

- The distributed heat release leads to low $NO_X$ emission, while CO emission remains low. Hence, the $NO_X$ advantage is not a result of a compromise. Compared to V-shaped flames, more than 50% lower $NO_X$ emission was achieved on average.

- There is no need for exhaust gas recirculation or oxidizer dilution, like in the case of MILD combustion.

- Flame luminosity is significantly lower than that of V-shaped and straight flames. Hence, optical sensing and control of the process bear further challenges if desired.

- The overall sound pressure level of distributed combustion was 6 dB lower than that of a V-shaped flame, meaning a notably reduced acoustic load on both the device and affected personnel.

- The time-averaged averaged acoustic spectrum contains well-localized peaks, which are related to the eigenmodes of the combustion system geometry. Consequently, its tendency to thermoacoustic instabilities is hypothesized to be significantly lower than that of V-shaped flames and similar to that of flameless and MILD combustion.

**Acknowledgments**

This paper is dedicated to the memory of Prof. Mário Costa. The help of Gyöngyvér Hidegh and Attila Kun-Balog in building the combustion test rig and assisting the measurements is greatly acknowledged.

**Funding**

The research reported in this paper was supported by the National Research, Development and Innovation Fund of Hungary, project №. OTKA-FK 124704 and TKP2020 NC, Grant No. BME-NC, based on the charter of bolster issued by the NRDI Office under the auspices of the Ministry for Innovation and Technology, and the János Bolyai Research Scholarship of the Hungarian Academy of Sciences.

**Conflict of interest**

The author declares that he has no known competing financial interests or personal relationships that could have appeared to influence the work reported in this paper.